\documentclass[prl,reprint,amsmath,amssymb,preprintnumbers,superscriptaddress,nofootinbib]{revtex4-1}

\usepackage{epsfig,epstopdf}
\usepackage{subfigure}
\usepackage{color}

\def\beq{\begin{equation}}
\def\eeq{\end{equation}}
\def\bea{\begin{eqnarray}}
\def\eea{\end{eqnarray}}

\newcommand{\lsim}{
\mathrel{\hbox{\rlap{\hbox{\lower4pt\hbox{$\sim$}}}\hbox{$<$}}}}
\newcommand{\gsim}{
\mathrel{\hbox{\rlap{\hbox{\lower4pt\hbox{$\sim$}}}\hbox{$>$}}}}

\newcommand{\dis}[1]{\begin{equation}\begin{split}#1\end{split}\end{equation}}
\newcommand{\mpl}{M_{\rm Pl}}

\begin{document}

\title{Implication of the swampland distance conjecture on the Cohen-Kaplan-Nelson bound in de Sitter space}

\author{Min-Seok Seo}
\email{minseokseo57@gmail.com}
\affiliation{Department of Physics Education, Korea National University of Education,
\\ 
Cheongju 28173, Republic of Korea}

\begin{abstract}
\noindent 

 The Cohen-Kaplan-Nelson (CKN) bound formulates the condition that  black hole is not produced by the  low energy effective field theory dynamics.
 In de Sitter space it also constrains the maximal size of the matter distribution to be smaller than the cosmological horizon determined by black hole.
 On the other hand, the swampland distance conjecture (SDC) predicts that  de Sitter space becomes unstable by the descent of the low energy degrees of freedom from UV.
 This results in the rapid increase in the energy inside the cosmological horizon, the distribution of which can be constrained by the CKN bound.
We study the CKN bound in de Sitter space in detail and point out that when compared with the slow-roll in the inflation, the bound on the slow-roll parameter which forbids the eternal inflation is obtained.

\end{abstract}
\maketitle

\section{Introduction}

 The swampland program \cite{Vafa:2005ui} (for  reviews, see  \cite{Brennan:2017rbf, Palti:2019pca}) aims at finding out  bounds on the low energy effective field theory (EFT) parameter space consistent with quantum gravity.  
 The swampland distance conjecture (SDC) \cite{Ooguri:2006in} provides the string-inspired explanation to these bounds by assuming that  all parameters  are determined by the dynamical stabilization of the moduli.
 It claims that as we move away from any point in the allowed parameter region to the boundary,   some modulus  traverses trans-Planckian geodesic distance and an infinite tower of states  becomes exponentially light.
Then the `species cutoff'  \cite{Dvali:2007hz, Dvali:2007wp}, above which gravity is no longer weakly coupled to matter, falls off drastically, which signals the breakdown of the EFT.

 Combined with the covariant entropy bound \cite{Bousso:1999xy}, the SDC is used to support the de Sitter (dS) swampland conjecture (dSC) \cite{Obied:2018sgi} (see also \cite{Andriot:2018mav, Garg:2018reu, Ooguri:2018wrx} for a refined version) which claims the quantum instability of dS space.
  The  increasing number of  excitations of light degrees of freedom along the trajectory of the inflaton, the modulus responsible for the vacuum energy,  results in the rapid increase in the entropy.
 Once this new entropy contribution exceeds the covariant entropy bound given by the horizon area, the EFT defined below the almost constant cosmological horizon scale $H$ is no longer valid \cite{Ooguri:2018wrx} (for relevant discussions, see, e.g., \cite{Seo:2018abc, Seo:2019mfk}).
From this we can estimate the duration of the quasi-dS phase \cite{Seo:2019wsh, Cai:2019dzj}, which is similar to that suggested by the trans-Planckian censorship conjecture \cite{Bedroya:2019snp} (see also \cite{Brahma:2020zpk}) but enhanced by the slow-roll parameter. 

 The argument above illustrates how the  excitations of a large number of states deform the cosmological horizon in thermodynamic language.
 Meanwhile, such a rapid production of matter  also enhances the energy inside the cosmological horizon and furthermore, may result in the formation of black hole, in which case quantum gravity is no longer irrelevant to the particle interactions.
 The similar issue was considered by Cohen, Kaplan, and Nelson (CKN) \cite{Cohen:1998zx}.
 They pointed out that for the  EFT defined in the region of the size $L$ to be valid, the UV cutoff should be low enough not to produce black hole through the EFT dynamics.
 It also reads that matter should not be concentrated within the radius of black hole having the same amount of energy. 
 In   flat space, this condition is written as
 \dis{L^3 \Lambda_{\rm UV}^4 \lesssim L \mpl^2,\quad {\rm or}\quad \Lambda_{\rm UV} \lesssim \Big(\frac{\mpl}{L}\Big)^{1/2}. \label{eq:CKN}}
 Comparing with the Bekenstein entropy bound \cite{Bekenstein:1980jp},
 \dis{L^3 \Lambda_{\rm UV}^3 \lesssim S_{\rm BH}=\pi L^2 \mpl^2 
 ,\quad {\rm or}\quad \Lambda_{\rm UV} \lesssim \Big(\frac{\mpl^2}{L}\Big)^{1/3},}
 the CKN bound provides the more stringent bound on $ \Lambda_{\rm UV}$.
 
 We can also impose that black hole should not be produced in the dS background for the validity of the EFT description.
 This dS CKN bound was formulated in \cite{Banks:2019arz}, which also constrains the maximal size of the matter distribution to be smaller than the cosmological horizon determined by the backreaction of black hole.
 We expect from the SDC that as the inflaton moves in quasi-dS space, a large amount of matter is produced inside the cosmological horizon.
 Then the maximal size of the matter distribution is identified with the deformed cosmological horizon.
 The dS CKN bound in \cite{Banks:2019arz} requires that this deformed horizon is smaller than the cosmological horizon determined by black hole of the same energy.
 In this article, we investigate the physical implications of this SDC-assisted dS CKN bound.

 The plan of this article is as follows.
 After summarizing essential features of the SDC, we formulate the dS CKN bound in \cite{Banks:2019arz} more concretely.
 Then we focus on the specific limit $GMH \ll 1$, where $M$ is interpreted as a total energy, to describe the contraction of the cosmological horizon at the early stage of the matter production analytically.
 By combining the SDC with the dS CKN bound, we provide the EFT validity condition and compared this with the slow-roll during the inflation.
As we will see, our condition forbids the eternal inflation.

 \section{Swampland distance conjecture}
 \label{Sec:SDC}
  
 Before considering the CKN bound, we briefly review aspects of the SDC relevant to our discussion.
 The  universality of the gravitational interaction  implies that given the UV cutoff scale $\Lambda_{\rm UV}$ and ${\cal N}$ particle species of the EFT, the strength of the gravitational interaction is given by $\alpha_{\rm grav}={\cal N}(\Lambda_{\rm UV}/\mpl)^2$.
 Thus, if $\alpha_{\rm grav}$ is ${\cal O}(1)$, the EFT weakly coupled to gravity is no longer valid, which sets the species cutoff $\Lambda_{\rm sp}=\mpl/\sqrt{{\cal N}}$.
 
 On the other hand, the SDC assumes the existence of a tower of states, the mass gap  of which depends on some modulus $\varphi$ as $\Delta m=\mpl{\rm exp}[-\lambda\Delta\varphi]$, where $\Delta \varphi$ is the geodesic distance traversed by $\varphi$ in Planck unit and $\lambda$ is a model dependent ${\cal O}(1)$ constant.
 Taking an ansatz ${\cal N}=(\mpl/\Delta m)^\alpha$ for the number of  particle species below $ \Lambda_{\rm sp}$, the requirement that the heaviest particle mass in EFT is smaller than $\Lambda_{\rm sp}$,
 \dis{{\cal N} \Delta m \lesssim \Lambda_{\rm sp}=\frac{\mpl}{\sqrt{\cal N}}}
 gives $\alpha=2/3$, or \cite{Grimm:2018ohb} 
 \footnote{See also \cite{Hebecker:2018vxz} for the earlier derivation, which also discussed the `loophole' of the entropy argument in \cite{Ooguri:2018wrx}.}
 \dis{\Lambda_{\rm sp}\sim \mpl e^{-\frac{1}{3}\lambda\Delta\varphi},\quad\quad
 {\cal N} \sim  e^{\frac{2}{3}\lambda\Delta\varphi}.\label{Eq:distance}}
 
The semi-classical EFT on the (quasi-)dS background has a natural UV cutoff scale $H$ \cite{Cheung:2007st, Weinberg:2008hq} (see also \cite{Prokopec:2010be, Gong:2016qpq}). 
For this EFT to be valid against the SDC, $H \lesssim \Lambda_{\rm sp}$ must be satisfied, which reads
\dis{\Delta\varphi \lesssim \frac{3}{\lambda}\log\Big(\frac{\mpl}{H}\Big).\label{Eq:phibound}}
 This bound indeed is consistent with the entropy argument of \cite{Ooguri:2018wrx} which supports the dSC.
 Meanwhile, the low energy excitations can be seen as in thermal equilibrium with  the (quasi-)dS background, in which case the temperature is given by  Gibbons-Hawking temperature $T=H/(2\pi)$ \cite{Gibbons:1977mu}.
 The condition that their contribution to the entropy inside the horizon given by $S \sim {\cal N} T^3 H^{-3} \sim {\cal N}$ cannot exceed the covariant entropy bound $\pi \mpl^2/H^2$  gives the same bound on $\Delta\varphi$ as  \eqref{Eq:phibound}.
 
  We note that in the entropy argument above, we used ${\cal N}T^3\sim {\cal N}H^3$, rather than ${\cal N}\Lambda_{\rm sp}^3$ for the entropy density.
 For this estimation to be valid, the number density of states having the energy $M$ in the range from $H$ to $\Lambda_{\rm sp}$ needs to be Boltzmann-suppressed such that ${\cal N}{\rm exp}[-M/T]<1$.
 At the early stage of the inflaton excursion, i.e., $\Delta \varphi \ll 1$, this is trivially satisfied.
 As $\Delta \varphi$ gets close to the bound in \eqref{Eq:phibound}, ${\cal N}$ becomes exponentially large enough to overcome the Boltzmann-suppression factor, but then $\Lambda_{\rm sp}$ is close to $H$. 
 Thus, in any case, we can employ ${\cal N}H^3$ for the entropy density. 
 For example, the states having the energy $M\simeq \Lambda_{\rm sp}$ are not negligible if 
 \dis{\log{\cal N}-\frac{\Lambda_{\rm sp}}{T} &= \frac{2}{3}\lambda\Delta\varphi-\frac{2\pi \mpl}{H}e^{-\frac13\lambda\Delta\varphi}\label{Eq:Eden}}
 becomes positive.
 For the observational bound $H<10^{13-14}$GeV imposed by the recent Planck observation \cite{Akrami:2018odb}, this happens after $\Delta \varphi \simeq (0.9)[(3/\lambda)]\log(\mpl/H)$, at which $\Lambda_{\rm sp}$ is very close to $H$.
 For the same reason, we use $\sim{\cal N}H^4$ for the energy density throughout this article.

\section{dS CKN bound}
\label{Sec:CKN}

 The CKN bound \cite{Cohen:1998zx} stems from the requirement that black hole cannot be the low energy excitation, or equivalently, the total energy inside the given volume is sufficiently low that black hole is not produced.
 In this section, we apply this argument to the dS background.
 
 An uncharged, nonrotating black hole in the dS background is described by the Schwarzschild-dS geometry,
 \dis{&ds^2=-f(r)dt^2+\frac{1}{f(r)}dr^2+r^2 d\Omega_2^2,
 \\
 & f(r)=1-\frac{2 GM}{r}-H^2 r^2.}
 Since $f(r)$ becomes  negative infinity as $r\to 0$ and $r\to \infty$, one finds that $f(r)>0$ in which $t$ is along the timelike direction is satisfied only for specific range of $r$ and the specific parameter $GMH$.
 Indeed, one of three solutions to $f(r)=0$ is negative, which is not physical.
 When another two solutions $r_1,~r_2$ to $f(r)=0$ are real and positive  ($r_1 < r_2$), $f(r)$ is positive for $r_1<r<r_2$ then $r_1$ and $r_2$ are interpreted as the black hole horizon and the cosmological horizon, respectively.

 \begin{figure}[!ht]
  \begin{center}
   \includegraphics[width=0.40\textwidth]{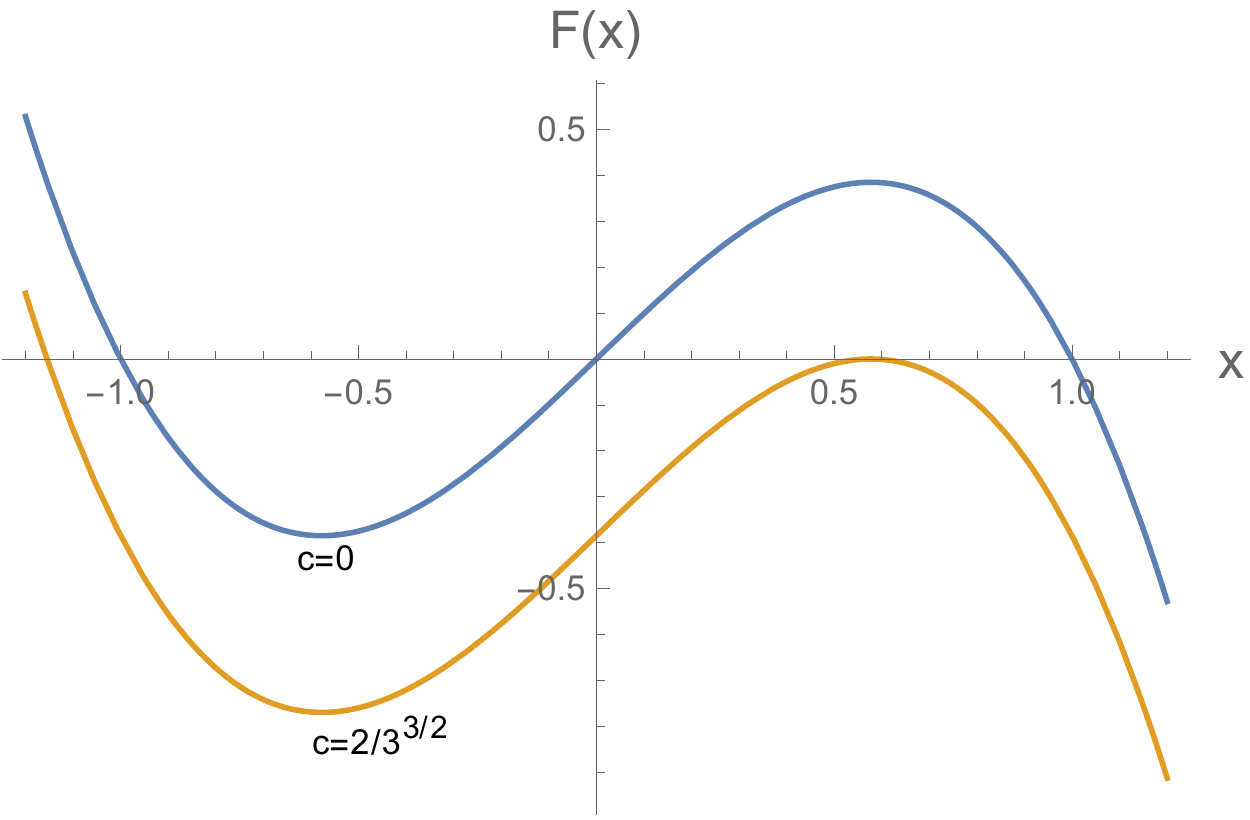}
  \end{center}
\caption{The behavior of $F(x)=c-x+x^3$ for two extreme cases, $c=0$ and $c=2/3^{3/2}$.
For $0 < c < 2/3^{3/2}$, $F(x)$ lies between two curves and both the black hole horizon and the cosmological horizon are well defined.
We note that for any value of $c$, $F(x)$ has a local maximum at $x=1/\sqrt3$.
  }
\label{Fig:F(x)}
\end{figure}
  
 From this observation, \cite{Banks:2019arz} suggested the dS CKN bound   by simply requiring $f(r)>0$.
Here we investigate this bound more quantitatively and discuss what happens when the SDC is additionally taken into account.
 For this purpose, we use the dimensionless parameter $x=H r$, such that $f(r)$ is rewritten as
 \dis{f(r)=-\frac{1}{x}\Big(2 GM H-x + x^3\Big)\equiv \frac{F(x)}{x},}
 thus $f(r)>0$ for $F(x)>0$.
 As depicted in Fig. \ref{Fig:F(x)}, we find that $F(x)=0$, or equivalently, $f(r)=0$ has two positive solutions for $0< 2GM < (2/3^{3/2})H^{-1}$.
 When $2GM = (2/3^{3/2})H^{-1}$, $F(x)$ vanishes at a local maximum $x=1/\sqrt3$ so two real solutions become the same.
 This shows that as the mass of black hole  increases, the cosmological horizon decreases until it coincides with the black hole horizon.
 Introducing a parameter $\theta$ defined as
 \dis{\cos\theta=-3^{3/2} GMH,\label{Eq:theta}}
 two horizons are given by
 \dis{H r_1=\sin\frac{\theta}{3}-\frac{1}{\sqrt3}\cos\frac{\theta}{3},\quad\quad &({\rm black~hole})
 \\
 H r_2=\frac{2}{\sqrt3}\cos\frac{\theta}{3}\quad\quad &({\rm cosmological}).\label{Eq:horizons}}
 We note that as $\theta$ increases over the region $\pi/2<\theta<\pi$,  which is equivalent to the increase in $2GM$ from $0$ to $(2/3^{3/2})H^{-1}$, $r_1$ monotonically increases from $0$ to $(1/\sqrt3)H^{-1}$ whereas $r_2$ monotonically decreases from $H^{-1}$ to $(1/\sqrt3)H^{-1}$.
 That is, $r_2$ is always larger than $r_1$.
 In perfect dS, in which the cosmological horizon size is just $H^{-1}$, the vacuum energy is the only contribution to the energy of spacetime.
 Since matter does not exist, black hole is not produced as $r_1=0$ indicates.
 When we add matter to this background, its backreaction results in the deformation of the cosmological horizon down to some value smaller than $H^{-1}$ and the geometry is no longer perfect dS.
 If matter forms black hole, the new cosmological horizon size will be $r_2$.
 
 For the EFT for matter in (quasi-)dS background to be valid without producing black hole, matter is required to be distributed over the region of the size $L$ larger than $r_1$, the radius of black hole whose  mass is  given by the corresponding energy. 
 The conjecture in \cite{Banks:2019arz} quantifies this no-black-hole-formation condition by  $f(L)>0$, or
 \dis{r_1 (\theta) < L < r_2 (\theta),\label{Eq:dSCKN}}
 where the energy is contained in $\theta$ through \eqref{Eq:theta}.
 We note that  $r_2$, the cosmological horizon determined by black hole, behaves like the upper bound of the  cosmological horizon for a fixed $M$. 
This may be justified from the fact that given $M$, black hole has the highest energy concentration.
Then we expect that the components of the energy-momentum tensor decay quickly as we move away from the black hole horizon and their values around the cosmological horizon are small compared to those of other form of the matter distribution.
This results in  the small local backreaction on the horizon.
\footnote{Unlike the black hole horizon, the (quasi-)dS cosmological horizon (in the absence of black hole) is an observer dependent object.
Even in this case, if the (quasi-)dS geometry is deformed, any observer must find the local backreaction of the energy-momentum tensor on the horizon (see, e.g., \cite{Gong:2020mbn}).
Of course, in the presence of black hole, the cosmological horizon becomes  observer-independent, as it is defined by the distance from the black hole center.}
 Another way to see this is to consider the case of $GMH \ll 1$, in which $r_2$ is very close to $H^{-1}$.
Suppose that matter is distributed almost homogeneously inside the horizon.
 If such a homogeneous distribution extends beyond $r_2$ to reach $H^{-1}$, the energy density of matter is indistinguishable from that of the vacuum energy.
 Then the region inside $H^{-1}$ keeps expanding exponentially.
 This is contradict to our expectation, the contraction of the cosmological horizon by the gravitational attraction of matter.
 If the fluctuation of the matter distribution is sufficiently small, in the absence of the dimensionful parameter comparable to $H$, the mean energy is the same   as that in the homogeneous case up to some constant, giving the similar result.
 This indeed is the way to treat the cosmological perturbations over the almost homogeneous and isotropic mean energy distribution.
 This issue will be visited in the next section, in which we find that the condition $L<r_2$ excludes the eternally inflating spacetime, supporting the argument above.

 \section{Deformation of dS space}
 \label{Sec:deformation}

  As we discussed in the previous section, the dS CKN bound describes the deformation (more precisely, contraction) of the cosmological horizon without invalidating the EFT in the presence of matter.
  In this section, we combine the dS CKN bound given by \eqref{Eq:dSCKN} with the SDC such that the energy inside the cosmological horizon  is increased by the increasing number of the particle species ${\cal N}$ along the inflaton trajectory.
  Here we use the energy density given by ${\cal N}T^4 \sim {\cal N}H^4$ as justified in the discussion around \eqref{Eq:Eden}.
  
  For the clear interpretation, we take the limit $GMH \ll 1$ of \eqref{Eq:dSCKN}.
  This obviously includes the flat space limit in which the cosmological horizon is much larger than the black hole size, but also describes the small deformation of the cosmological horizon by the small amount of matter.
  Expanding $\theta$ defined in \eqref{Eq:theta} around $\pi/2$, one finds that
  \dis{r_1\simeq 2GM,\quad\quad r_2\simeq H^{-1}-GM}
 for $GMH \ll 1$.
 Therefore, the dS CKN bound \eqref{Eq:dSCKN} reads
 \dis{GM < {\rm min}\Big[\frac{L}{2}, \frac{1}{H}-L\Big].\label{Eq:Mbound}}
 This shows that $GM$ is bounded by $L/2$ for $L<(2/3)H^{-1}$ which includes the flat space limit and by $H^{-1}-L$ for $L>(2/3)H^{-1}$, which describes the small deformation of the cosmological horizon, respectively.
 \footnote{ The bound \eqref{Eq:Mbound} tells us that if we set $L^{-1}$, the bound \eqref{Eq:Mbound} only allows for $M=0$, excluding any finite mass.
  This may imply that $L$ in fact cannot be arbitrarily fixed, but determined by the backreaction from matter in dS.
  In the absence of any matter, i.e., $M=0$, the horizon does not change at all, so cosmological size is just given by $H^{-1}$.
  This motivates us to interpret the CKN bound as a horizon size deformed by the matter in dS.}
 In any case, the bound on $GM$ is smaller than $(3H)^{-1}$, hence by setting $M={\cal N}H^4 L^{3}$ we expect the naive bound
 \footnote{While we take the mass density ${\cal N}H^4$ for the radiation, one may consider the mass density ${\cal N}H^2\mpl^2$ motivated by   the energy density of $\varphi$, the dark energy.
 The latter gives more stringent bound ${\cal N}\leq 1$, restricting $\Delta \varphi$ to the value close to zero.}
 \dis{{\cal N} \lesssim \frac{\mpl^2}{3 H^5 L^3}.}
 Then for $(2/3)H^{-1} \ll L\lesssim H$, i.e., when $L$ is very close to the dS horizon,  we obtain the bound \eqref{Eq:phibound} with the help of \eqref{Eq:distance}.
 However, this estimation lacks the shrinking bound on $M$ in the perfect dS limit $L \simeq H^{-1}$, in which case the upper bound  on $M$ vanishes.
 Therefore, we concentrate on the explicit bound given by \eqref{Eq:Mbound}.

 Before discussing the deformation of the cosmological horizon, we briefly address the case of $L<(2/3)H^{-1}$.
 This  includes the flat space limit, in which the energy $M={\cal N}H^4 L^{3}$ is concentrated on the small region deep inside the cosmological horizon.
 Then the bound $GM<L/2$ gives
 \dis{{\cal N}\lesssim \frac{\mpl^2}{H^4 L^2},}
 which is trivially satisfied for $L<(2/3)H^{-1}$ since the SDC imposes 
 \dis{{\cal N}\sim e^{\frac23\lambda\Delta\varphi} \lesssim \Big(\frac{\mpl}{H}\Big)^2.}

 We now move onto the case of $L>(2/3)H^{-1}$, in which $L$ is very close to $H^{-1}$.
 The dS CKN bound in this case can be combined with the SDC by considering the  matter production in quasi-dS space induced by the  excursion of $\varphi$ as well as the interaction between matter and the background.
 As $\varphi$ moves, ${\cal N}$ is drastically increased by the descent of a tower of states.
 As the backreaction from the produced matter on the background  deforms the cosmological horizon, matter is distributed inside the new cosmological horizon of the size $L$ which is slightly smaller than $H^{-1}$.
 The dS CKN bound  imposes that $L$ is  smaller than $r_2$, the change in the size of  cosmological horizon when the produced matter forms black hole.

 To see this more explicitly, we note that  time and length scales in our discussion such as the constant dS cosmological horizon size $H^{-1}$  are written in terms of the static coordinates.
 Meanwhile, for the comparison with the inflationary cosmology in which the slow-roll parameters and the matter fluctuation are defined in terms of the flat coordinates $(t, r)$, we need the relation between two coordinates, 
  \dis{&e^{-Ht_s}=e^{-Ht}\sqrt{1-H^2 r_s^2},
 \\
 &r_s=r e^{Ht}.\label{eq:statflat}}
 From this,  we find that at fixed $r_s$, say, $r_s=H^{-1}(1-\gamma)$ with a constant $\gamma$, we have $t_s=t-\log(2\gamma)/2$ thus $dt_s=dt$.
 This indicates that so far as we are interested in the fixed $r_s$, $dt$ and $dt_s$ can be used interchangeably.
 Moreover,   as can be inferred from the fact that the time evolution of  $\varphi$ breaks the dS isometry associated with the time translation, the quantum fluctuation of the inflaton $\varphi$ is identified with the fluctuation of the flat time coordinate, $\Delta t$.
 Under such fluctuation, the energy density of matter inside the horizon $\rho={\cal N}H^4$ fluctuates to give $\Delta\rho =\rho [\Delta(e^{Ht})/e^{Ht}]=({\cal N}H^4)\times H\Delta t$.
 Now  suppose we are in the empty  dS space with the almost constant horizon size $H^{-1}$.
  From this, we separate the dS background from matter produced by the density fluctuation,  to quantify how the backreaction from matter deforms the background geometry.
  Then for the given energy density $\rho={\cal N}H^4$, the mass produced through the fluctuation becomes
 \dis{M={\cal N}H^4 (H\Delta t) (\frac{4\pi}{3} H^{-3})= \frac{4\pi}{3}{\cal N}H^2\Delta t.\label{eq:Mtotal}}
 If $H$ is constant, the fluctuation leading to $\Delta t$ results in the increase of $M$ inside the horizon as ${\cal N}$ increases by the SDC, the backreaction of which leads to the shrink of the cosmological horizon by $H^{-1}-L$.

 Meanwhile, in the inflationary cosmology, $\varphi$ not just increases ${\cal N}$ through the fluctuation giving $\Delta t$, but also slow-rolls on the potential, resulting in the decrease of $H$ in time.
 Then two effects that change the size of the cosmological horizon compete : the shrink by $H^{-1}-L$ through the matter production, and the growth through the slow-roll of $\varphi$, which is parametrized by the slow-roll parameter.
 If the former dominates over the latter, inflation never ends, realizing the eternal inflation.
 From the expression for the slow roll parameter $\epsilon$,
 \dis{\epsilon=-\frac{\dot H}{H^2}=\frac{4\pi {\dot \varphi}^2}{H^2},\label{Eq:epsilon}}
 one obtains 
 \dis{\frac{d H}{d\varphi}=-H\sqrt{4\pi \epsilon}.}
 Treating $\epsilon$ as almost constant, the increase in the horizon size induced by the change in $\varphi$ is given by
 \dis{\Delta H^{-1}_\varphi = H^{-1}\sqrt{4\pi \epsilon}\Delta\varphi  = \epsilon  \Delta t \label{eq:newL}}
 where for the last equality, the relation $\Delta\varphi={\dot\varphi}\Delta t=[\epsilon/(4\pi)]^{1/2}H\Delta t$ is used (see \eqref{Eq:epsilon}).

 Given the total mass \eqref{eq:Mtotal}, the shrink of the horizon $H^{-1}-L$ is at most $GM$, as constrained by the dS CKN bound $GM<H^{-1}-L$.
For the inflation to end, such a shrink  must be compensated by the increase in the horizon size by the slow-roll given by \eqref{eq:newL}, giving $\Delta H^{-1}_\varphi > GM$, or
 \dis{{\cal N}\lesssim  \sqrt{4\pi \epsilon}\frac{\mpl^2}{H^3}\frac{\Delta\varphi}{\Delta t}=\epsilon\frac{\mpl^2}{H^2}.\label{Eq:dSCKNdef}}  
 To see the physical meaning of this   bound,  we assume  $\epsilon \ll 1$ such that $L$ is still close to $H^{-1}$ after a single $e$-fold, $H\Delta t \sim 1$.
 This means that $t\sim H^{-1}$ can be regarded as the early stage of the matter production, i.e., $\Delta \varphi \ll 1$ thus UV degrees of freedom do not descend below $H$ yet and ${\cal N}\sim {\cal O}(1)$.
 From ${\cal N}\simeq 1+(2/3)\lambda \Delta \varphi+\cdots$ one finds that \eqref{Eq:dSCKNdef} provides the bound on $\epsilon$,
 \footnote{While an expansion of ${\cal N}$ contains ${\cal O}(\epsilon)$ term given by $(2/9)\lambda^2(\Delta\varphi)^2$, a coefficient of $\epsilon$ is ${\cal O}(1)$, which is much smaller than $\mpl^2/H^2$ in \eqref{Eq:dSCKNdef}. 
 Thus, this term does not affect the result significantly.}
 \dis{\epsilon \gtrsim \frac{H^2}{\mpl^2}+\frac{\lambda}{6\sqrt{\pi}}\frac{H^3}{\mpl^3}+\cdots.}
 Here  the $\lambda$ dependence is suppressed as $H\ll \mpl$ and we are left with the model-independent inequality $\epsilon \gtrsim (H/\mpl)^2$.
  This bound  is  nothing more than the well-known condition that the eternal inflation does not take place \cite{Steinhardt:1982, Vilenkin:1983xq, Linde:1986fc, Linde:1986fd, Goncharov:1987ir} (see also \cite{Starobinsky:1986fx}).
 \footnote{The conventional condition $\epsilon \gtrsim (H/\mpl)^2$ prohibiting the eternal inflation comes from the width of the probability distribution for the deviation of $\varphi$ from the classical trajectory.
 Since the exact coefficient attached to $(H/\mpl)^2$  is obtained from the Fokker-Planck equation, which may be model dependent, we just consider the parametric dependence, by assuming ${\cal N}_0$ is not too large.
 For recent discussions along the line of this article are given in, for example, \cite{ArkaniHamed:2007ky, Kinney:2018kew, Brahma:2019iyy, Rudelius:2019cfh}. 
 If the universe is in the quantum mechanically excited state, the eternal inflation may take place for $\epsilon$ larger than  the integer multiples of $H^2/\mpl^2$ \cite{Seo:2020ger}.}
 The eternal inflation occurs when the cosmological fluctuations, the wavelength of which is stretched beyond the cosmological horizon (thus the low energy degrees of freedom below $H$), enhance the probability that the inflaton  does not roll down its potential.
Recalling the discussion at the end of previous section, one finds   the dS CKN bound $L< r_2$ imposes that the descent of states claimed by the SDC induces the contraction of the horizon by their gravitational attraction, so it is natural that the eternal inflation is not favored.

 As $\varphi$ keeps moving on, the number of low energy degrees of freedom exponentially increases in time.
 Since $\Delta t$ is no longer small, the matter produced fills the region inside the horizon almost homogeneously, which is in thermal equilibrium with the background.
 This results in the increase in $M$ toward $(3^{3/2} GH)^{-1}$ such that  $\theta$ gets close to $\pi$.
  Around $\theta= \pi$,  the   matter distribution must satisfy both $L \simeq (1/\sqrt3)H^{-1}$  and $GM=(1/3^{3/2})H^{-1}$.
From $M={\cal N}H^4 L^3 \simeq {\cal N}H$ we find that ${\cal N}\simeq (\mpl/H)^2$, which is close to the saturation of the bound \eqref{Eq:phibound}, and \cite{Ooguri:2018wrx} claims that  $\epsilon$ gradually increases and eventually becomes ${\cal O}(1)$.
More concretely,  once the bound ${\cal N}\leq (\mpl/H)^2$ is saturated, the horizon is strongly deformed not to violate the bound. 
The shrink of the horizon is forbidden by the rapid increase in $H^{-1}$, as $\epsilon \sim {\cal O}(1)$ implies. 
This is achieved when the rate of change in ${\cal N}$ is still smaller than that of $(\mpl/H)^2$, i.e., $(d{\cal N}/d\varphi){\dot \varphi}\lesssim (-2{\dot H}/H^3)\mpl^2$.
Since $d{\cal N}/d\varphi=(2/3)\lambda{\cal N}=(2/3)\lambda (\mpl/H)^2$ at the saturation, from \eqref{Eq:epsilon} we obtain  (see also \cite{Seo:2019wsh})
\dis{\epsilon \gtrsim \frac{\lambda^2}{36\pi}.}
As estimated in \cite{Seo:2019wsh, Cai:2019dzj}, the saturation takes place after about ${\cal N}\sim (1/\langle \epsilon\rangle)^{1/2} \log(\mpl/H)$ $e$-folds, where $\langle\epsilon \rangle$ is the averaged value of $\epsilon$ in the quasi-dS regime hence smaller than unity.
Since $H/\mpl <10^{-5}$, if $\langle\epsilon \rangle$ is as small as $10^{-2}$, $\varphi$ slow-rolls over $60~e$-folds, which is sufficient to resolve the horizon problem or the flatness problem \cite{Seo:2019wsh}.


Furthermore, the refined dSC \cite{Andriot:2018mav, Garg:2018reu, Ooguri:2018wrx} imposes another condition   $\dot{\epsilon}/(H\epsilon)\sim {\cal O}(1)$ for $\epsilon \ll 1$.
If it were the case, even if $\epsilon \simeq 0$ at some moment, it soon increases as $\varphi$ moves.  
This in fact can be frequently found in the supergravity models of the inflation, which is known as the eta problem \cite{Copeland:1994vg}.

 \section{Conclusion }
  
 So far, we have discussed the implications of the CKN bound when we take the SDC into account.
 In  quasi-dS space, the exponential descent of the low energy degrees of freedom as conjectured  by the SDC leads to the rapid increase in the energy inside the cosmological horizon.
 The dS CKN bound states that for the EFT not to be spoiled by the production of black hole,  matter must be distributed between the black hole radius and the cosmological horizon determined by  black hole of the same energy.
 By identifying the maximal size of the matter distribution with the cosmological horizon deformed by the production of matter, we find that the condition for the SDC-assisted dS CKN bound to forbid the eternal inflation coincides with the well-known condition on the slow-roll parameter, $\epsilon \gtrsim H^2/\mpl^2$.
 Our discussion is supplementary to the entropy argument supporting the dSC, as it enables us to investigate more detailed dynamics of the contracting cosmological horizon.
 
 Meanwhile, the SDC seems to be consistent with other swampland conjectures, such as the weak gravity conjecture (WGC) \cite{ArkaniHamed:2006dz}.
 The weak gravity conjecture  quantifies the quantum gravity obstruction to restoring a global symmetry by taking vanishing limit of the unbroken U(1) gauge coupling (for a recent  discussion  relevant to the CKN bound, see, e.g., \cite{Davoudiasl:2021aih} and for discussions on the WGC in dS space, see, e.g., \cite{Huang:2006hc, Montero:2019ekk, Montero:2021otb}).
  The correlation between the WGC and the SDC can be found from, for example, the restoration of the global shift symmetry in the limit of infinite field value of the modulus, where non-perturbative effects become suppressed  \cite{Palti:2017elp, Grimm:2018ohb, Lee:2018urn, Lee:2018spm, Gendler:2020dfp}.
  The SDC claims that this can be prevented by cutting off $\Delta \varphi$ at the Planckian value. 
  The quantum gravity implications of the connection between the SDC, the dSC, the weak gravity conjecture and the CKN bound  will be a subject of further investigation.

\vspace{5mm}
\begin{acknowledgments}


\end{acknowledgments}


\end{document}